\documentclass[letter,scriptaddress,twocolumn, prl,showkeys]{revtex4}
\usepackage{amssymb}
\usepackage{amsmath}
\usepackage{makeidx}
\usepackage{amsfonts}
\usepackage{graphicx,epstopdf}
\usepackage[ansinew]{inputenc}
\usepackage[usenames,dvipsnames]{pstricks}
\usepackage{subfigure}
\usepackage{pdfpages}
\usepackage{epsfig}
\usepackage{pst-grad}
\usepackage{pst-plot}
\usepackage[colorlinks,hyperindex]{hyperref}
\usepackage[dvipsnames]{xcolor}

\hypersetup{
colorlinks,
citecolor=blue,
linkcolor=red,
urlcolor=red,
}

\makeindex

\begin{document}

\title{Modeling Dark Matter Halos with Nonlinear Field Theories}
\author{R. A. C. Correa$^{1}$\footnote{fis04132@gmail.com}, P. H. R. S. Moraes$^{2}$\footnote{%
moraes.phrs@gmail.com}, A. de Souza Dutra$^{3}$\footnote{asdutra@gmail.com}, O. L. Dors$^{4}$\footnote{%
olidors@univap.br}, W. de Paula$^{1}$\footnote{%
wayne@ita.br}, and T. Frederico$^{1}$\footnote{%
tobias@ita.br}}

\vspace{1cm}

\affiliation{$^{1}$ Instituto Tecnol\'ogico de Aeron\'autica, DCTA, 12228-900, S\~ao Jos\'e dos
Campos, SP, Brazil\\
$^{2}$ Instituto de Astronomia, Geof\'isica e Ci\^encias Atmosf\'ericas, Universidade de S\~ao Paulo, 05508-090, S\~ao Paulo, Brazil\\
$^{3}$ Universidade Estadual Paulista-Campus de Guaratinguet\'a, 12516-410, Guaratinguet\'a, SP, Brazil\\
$^{4}$ Universidade do Vale do Para\'iba, 12244-000, S\~ao Jos\'e dos Campos, SP,Brazil}

\begin{abstract}
In the present work, we adopt  a nonlinear scalar field theory coupled to the gravity sector  to model galactic dark matter.  We found analytical solutions for the scalar field coupled to gravity in the Newtonian limit, assuming an isotropic spacetime and a field potential, with a  position dependent form  of the superpotential, which entails the nonlinear dynamics of the model with self-interactions. The model introduces a position dependent 
enhancement of the self-interaction of the scalar fields towards the galaxy center, and while going towards the  galaxy border  the interaction tends to vanish building a non self-interacting DM scenario.  The developed approach is able to provide  a reasonable analytical  description of the rotation curves in both dwarf and low surface brightness late-type galaxies, with parameters associated with the dynamics of the scalar field.
\end{abstract}

\date{July 2, 2020}
\maketitle

\section{1. Introduction}

Dark Matter (DM) is one of the most important open problems in Physics.
It does not seem to interact with electromagnetic force and, therefore, it cannot be directly seen.  However, its gravitational effects are essential to explain the structure formation and 
mass distribution of galaxies.

The existence of DM is well established
by several experimental cosmological observations (e.g. see \cite{tanabashi18, mcGaugh20} for a review) and it
is necessary, for instance, to explain the spiral galaxy rotation curves.  While the classical newtonian gravity theory
requires that the {\it orbital circular velocity vs.  distance to galactic center} curve after attaining its maximum  decreases as one moves away from the galactic center,  observations carried out along decades have found that the velocity remains approximately constant in this interval \cite{freman70,rubin/1982,rohlfs/1986, begeman91, mcGaugh16}. In order to explain such a phenomenon, a DM halo is supposed to exist and to be responsible for most of the galaxies mass \cite{navarro/1996,jenkins/2001,dubinski/1991}. Moreover, 
DM has also a fundamental role in the large-scale structure formation in the universe \cite{blumenthal/1984,davis/1985}.

Although the gravitational effects of DM are notorious, despite several efforts, no particle associated with DM has ever been detected \cite{liu/2017,bi/2013}. This fact has led some theoretical physicists to claim that DM does not exist and its observational effects are due to some geometrical correction terms to General Relativity. From this perspective, it was shown that it is indeed possible to describe structure formation \cite{dodelson/2006,acquaviva/2005,bebronne/2007,koyama/2006,pal/2006} and rotation curves of galaxies \cite{deliduman/2020,o'brien/2018,mak/2004,capozziello/2004} through extended theories of gravity. Here we will attain to the several observational evidences of DM existence, which besides the aforementioned cases, refers to gravitational lensing \cite{clowe/2004} and to the well-known Bullet Cluster \cite{clowe/2007}.

Nowadays, there is a plethora of DM particle candidates. For instance: {\bf i)} axions, which are hypothetical particles whose existence was postulated to solve the strong CP problem of quantum chromodynamics \cite{visinelli/2009}; {\bf ii)} sterile neutrinos, which interact only gravitationally with ordinary matter \cite{boyarsky/2009}; {\bf iii)} WIMPs (weakly interactive massive particles), which arise naturally from theories that seek to extend the standard model of particle physics, such as supersymmetry \cite{arcadi/2018}. In particular, although WIMPs are the most studied class of DM particle candidates, current DM direct and indirect detection experiments have not yet discovered compelling signals of them \cite{bauer/2015}. In fact, recent data from the Large Hadron Collider have found no evidence of a deviation from the standard model on GeV scales \cite{aad/2013}. It is evident that the microscopic nature of DM is sufficiently unsettled as to justify the consideration of alternative candidates.

Among these candidates one can quote the Bose-Einstein condensate (BEC) coupled to gravity. In this model, the nature of DM is completely determined by a fundamental scalar field endowed with a scalar potential  \cite{ji/1994,guzman/2000,guzman/1999,lee/1996,elisa1,elisa2,elisa3,elisa4}. In such a context, DM halos can be described as a condensate made up of ultra-light bosons.

In the next section we will introduce the mathematical framework in which DM is described by a scalar field. For now it is interesting to quote that BEC DM has shown to provide a good fit to the evolution of cosmological densities \cite{matos/2009} and acoustic peaks of the cosmic microwave background \cite{rodriguez-montoya/2010}. It has also been applied to the rotation curves of galaxies \cite{matos/2017,guzman/2015,kun/2020,craciun/2019}. The growth of perturbations in an expanding Newtonian universe with BEC DM was studied in \cite{chavanis/2012}. In \cite{jusufi/2019}, the possibility of wormhole formation in the galactic halo due to BEC DM was investigated. 

It is very important to remark that recently BEC was observed in the Cold Atom Laboratory \cite{aveline/2020}, which is orbiting Earth on board the International Space Station. The microgravity environment of such an experiment allowed for the observation of BEC during approximately 1s, instead of $\sim $ms, which is the case of ground experiments. The continuous and increasing observations of BEC will naturally allow for the understanding its properties and the viability to  represent DM on a galaxy environment. 

For our purposes in the present work it is fundamental to remark that some classes of important physical systems are intrinsically nonlinear, specially those systems that supports topological defects \cite{c1,c2,c3,c4}. Nonlinear structures play an important role in the development of several branches of physics, such as cosmology \cite{Vilenkin}, field theory \cite{Weinberg, Vachaspati}, condensed matter physics \cite{Bishop} and others \cite{Gu}. For example, we can find nonlinear configurations in various contexts, as the oscillons in the standard model-extension \cite{Gleiser,Rafael,Kaloian}, during the formation of the aforementioned BECs \cite{Tobias, Malomed}, in supersymmetric sigma models \cite{Butter}, in Yang-Mills theory \cite{Tanizaki} and in Lorentz breaking systems \cite{Dutra}.

Particularly, in a cosmological context, we know that nonlinear scalar field theories play a significant role in our understanding of the cosmological dynamics and structure formation. Both the inflationary epoch and the current phase of dark energy domination can be modeled using nonlinear scalar field models.  For example, in a cosmological scenario coming from multi-component scalar field models, it was shown in \cite{RafaelEPJC} that nonlinear interactions are responsible for providing a complete analytical cosmological scenario, which describes the inflationary, radiation, matter, and dark energy eras. Also within a cosmological context, recently, Adam and Varela \cite{c5} have introduced a very interesting concept of inflationary twin models, where generalized K-inflation theories can be controlled in a simple way, thereby allowing to describe the cosmological evolution during the inflation period.

On the subject of BEC DM in nonlinear models, it is well known that dark matter halos can be modeled with the so-called solitons \cite{Mielke, Schunck,li/2014,mielke/2004}. In this case, nonlinear configurations can provide a powerful description of the currently observed rotation curves. However, as a consequence of the nonlinearity, in several BEC DM models we lose the ability to obtain a complete set of analytical  solutions. Therefore, new insights and methods to solve BEC DM problems analytically in nonlinear backgrounds are a major challenge that needs to be considered in order to deepen  and enlarge our understanding of the physics brought by this framework.

The main goal of this work is to show an analytical approach that can be used in general to study BEC DM within  nonlinear scenarios. Our aim is to investigate possible galactic DM models based on nonlinear scalar field theories coupled to the gravity sector. The  validation of these models is obtained by comparing  predictions for galaxy rotation   curves  with observational data.

This work is organized as follows: in Sect. \textcolor{red}{2} we introduce the framework which will be studied here. In Sect. \textcolor{red}{3} we present our approach and analytical solutions. In Sect. \textcolor{red}{4} DM halos are analyzed. Finally, Sect. \textcolor{red}{5} provides our conclusions.

\section{2. Framework}

In this section, we introduce the scenario which will be studied in this work. Let us assume a theoretical framework where DM consists of a complex scalar field \cite{Mielke}, which is responsible for producing  galactic halos through the Bose-condensed state coupled to gravity. In this case, we can write the Einstein-Hilbert action in the following form \cite{Schunck}

\begin{equation}
\mathcal{S=}\int dx^{4}\sqrt{-g}\left[ \frac{R}{16\pi G}+\frac{1}{2}%
g^{\mu \nu }\partial _{\mu }\psi \partial _{\nu }\psi ^{\ast }-V(\psi \psi
^{\ast })\right] , \label{1}
\end{equation}
\noindent where $R$ is the curvature scalar, $G$ is the gravitational constant, $g$ corresponds to the determinant of the metric $g_{\mu \nu}$, $\psi$ is the complex scalar field and $V$ its potential. We are adopting units such that  $c=\hbar=1$. For the complex scalar field model, the 
DM density results from the difference between the number density of bosons and of their antiparticles \cite{li/2014}. Moreover, there are some reasons for considering a complex field rather than a real one. Among those is the U(1) symmetry corresponding to the DM particle number conservation \cite{boyle/2002}.

Since  $\psi$ is a complex scalar field, we will break $\psi$ up into two real fields, one associated with the real part and {another one} associated with the imaginary part:
\begin{equation}
    \psi(r,t)=\phi(r,t)+i \chi(r,t) \label{dc}.
\end{equation}

One can see that Equation~(\ref{1}) becomes
\begin{eqnarray}
&&\left. \mathcal{S}\mathcal{=}\int dx^{4}\sqrt{-g}\left[ \frac{R}{16\pi
G}+\frac{1}{2}g^{\mu \nu }(\partial _{\mu }\phi \partial _{\nu }\phi
+\partial _{\mu }\chi \partial _{\nu }\chi )\right. \right.   \nonumber \\
&&\left. -V(\phi ,\chi )\right] .  \label{1.1}
\end{eqnarray}

Now the scalar sector behaves like a two-field theory, where $\phi$ and $\chi$ are real fields. Therefore, from the principle of least action $\delta S=0$, we obtain both Einstein and motion equations for the system. Firstly, let us apply the variation of Eq.~(\ref{1.1}) in regard to the metric $g_{\mu \nu}$. In this case, we obtain the Einstein equation for the system
\begin{equation}
R_{\mu \nu }-\frac{1}{2}g_{\mu \nu }R=kT_{\mu \nu }\, .  \label{ae}
\end{equation}
\noindent Moreover, we are using the definition $\kappa = 8\pi G$ and $T_{\mu \nu}$ is the energy momentum tensor, represented by
\begin{equation}
T_{\mu \nu }=(\partial _{\mu }\phi \partial _{\nu }\phi
+\partial _{\mu }\chi\partial _{\nu }\chi )-g_{\mu \nu}\mathcal{L},  \label{4}
\end{equation}

\noindent where $\mathcal{L}$ represents the Lagrangian density of the scalar field, that reads
\begin{equation}
\mathcal{L}=\frac{1}{2}\left( g^{\mu \nu }\partial _{\mu }\phi
\partial _{\nu }\phi+ g^{\mu \nu }\partial _{\mu }\chi
\partial _{\nu }\chi\right)-V(\phi ,\chi ) .
\label{5}
\end{equation}

Secondly, applying the variation regarding the scalar fields, we have
\begin{eqnarray}
\frac{1}{\sqrt{-g}}\partial _{\mu }\left( \sqrt{-g}g^{\mu \nu }\partial
_{\nu }\phi \right) +\frac{\partial V}{\partial \phi } &=&0   \label{fe1} 
\end{eqnarray}
{and}
\begin{eqnarray}
\frac{1}{\sqrt{-g}}\partial _{\mu }\left( \sqrt{-g}g^{\mu \nu }\partial
_{\nu }\chi \right) +\frac{\partial V}{\partial \chi } &=&0.  \label{fe2}
\end{eqnarray}

Our purpose in this work is to derive analytical solutions for the equations of motion considering spherically symmetric system. 
Therefore,  we can write the line element as
\begin{equation}
ds^2 = e^{\alpha (r)} dt^2 - e^{\beta (r)} dr^2 - r^2 \left(
 d\theta^2 + \sin^2\theta d\varphi^2 \right), \label{met}
\end{equation}
{being  $\alpha$ and $\beta$ the metric potentials.} 

From the above metric, the stress-energy tensor (\ref{4}) becomes diagonal and is given by

\begin{equation}
T_{\mu }^{\mu }=\text{diag}(\rho ,-p_{r},-p_{\perp },-p_{\perp }),
\label{6}
\end{equation}

\noindent where $\rho$ is the energy density, $p_{r}$ and $p_{\perp }$ are the radial and tangential components of the pressure. Thus, using Eq.~(\ref{met}) into Eq.~(\ref{4}), we obtain

\begin{eqnarray}
&&\left. \rho =\frac{e^{-\alpha }}{2}\left[ (\partial _{t}\phi )^{2}+(\partial
_{t}\chi )^{2}\right] +\frac{e^{-\beta }}{2}\left[ (\partial _{r}\phi
)^{2}+(\partial _{r}\chi )^{2}\right] \right.   \nonumber \\
&&\left. +V(\phi ,\chi ),\right.   \label{ds1} \\
&&  \nonumber \\
&&\left. p_{r}=\frac{e^{-\alpha }}{2}\left[ (\partial _{t}\phi )^{2}+(\partial
_{t}\chi )^{2}\right] +\frac{e^{-\beta }}{2}\left[ (\partial _{r}\phi
)^{2}+(\partial _{r}\chi )^{2}\right] \right.   \nonumber \\
&&\left. -V(\phi ,\chi ),\right.   \label{ds2} \\
&&  \nonumber \\
&&\left. p_{\perp }=\frac{e^{-\alpha }}{2}\left[ (\partial _{t}\phi
)^{2}+(\partial _{t}\chi )^{2}\right] -\frac{e^{-\beta }}{2}\left[
(\partial _{r}\phi )^{2}+(\partial _{r}\chi )^{2}\right] \right.   \nonumber
\\
&&\left. -V(\phi ,\chi ),\right.   \label{ds3}
\end{eqnarray}

\noindent where we are using the notation $\partial_t\equiv \partial/\partial t$ and $\partial_r\equiv \partial/\partial r$.

After straightforward manipulations, the non-vanishing components of the Einstein equations can be put in the form
\begin{eqnarray}
\alpha' + \beta' & = & \kappa (\rho + p_{\rm r}) r e^\beta  \label{nula}
\end{eqnarray}
{and}
\begin{eqnarray}
\beta' & = & \kappa \rho r e^\beta - \frac {1}{r} (e^\beta - 1). \label{la}
\end{eqnarray}

Finally, using Eq.~(\ref{met}) into Eqs.~(\ref{fe1}) and (\ref{fe2}), we find the following equations of motion 
 
\begin{align}
e^{\beta -\alpha }\ddot{\phi}-\phi ^{\prime \prime }-\left( \frac{\alpha
^{\prime }-\beta ^{\prime }}{2}+\frac{2}{r}\right) \phi ^{\prime }+e^{\beta
}V_{\phi }& =0  \label{mov1} 
&   
\end{align}
{and}
\begin{align}
e^{\beta -\alpha }\ddot{\chi}-\chi ^{\prime \prime }-\left( \frac{\alpha
^{\prime }-\beta ^{\prime }}{2}+\frac{2}{r}\right) \chi ^{\prime }+e^{\beta
}V_{\chi }& =0,  \label{mov3}
\end{align}

\noindent where dot represents derivative with respect to $t$ and prime derivative with respect to $r$. Moreover, $V_\phi \equiv \partial V/ \partial \phi$ and $V_\chi \equiv \partial V/ \partial \chi$.

Our main goal in the next sections will be to generate analytical solutions for the  equations above in the Newtonian limit,  i.e. when we have low velocity, weak interaction and weak gravitational
field. Furthermore, we  also {propose}  a procedure which is general when applied to study scalar field DM in a spherically symmetric space-time.

\section{3. The Method}

In this section, in order to obtain analytical solutions for the system under analysis, we will demonstrate a general approach which allows to reduce the second-order differential
equations (\ref{mov1}) and (\ref{mov3}) to first-order ones, whose general solution can be constructed by means of standard methods.

A DM halo comprising a BEC has a relatively low mean mass density so that we can use the Newtonian approximation. In this limit, the metric potentials $\alpha$ and $\beta$ are constants $\ll1$ so that $e^\alpha \simeq e^\beta \simeq 1$. Then, the equations of motion (\ref{mov1}) and (\ref{mov3}) become

\begin{align}
\ddot{\phi}-\phi ^{\prime \prime }-\frac{2}{r}\phi ^{\prime }+V_{\phi }& =0
\label{mov5} 
\end{align}
{and}
\begin{align}
\ddot{\chi}-\chi ^{\prime \prime }-\frac{2}{r}\chi ^{\prime }+V_{\chi }& =0.
\label{mov6}
\end{align}

We will focus our analysis in static configurations, where $\phi=\phi(r)$ and $\chi=\chi(r)$. However, we {emphasize}  that, given the static solution, one can apply a Lorentz boost in order to obtain a moving solution. Therefore, using the above assumption into Eqs. (\ref{mov5}) and (\ref{mov6}), we obtain the following coupled second-order differential equations

\begin{align}
\phi ^{\prime \prime }+\frac{2}{r}\phi ^{\prime }& =V_{\phi }  \label{mov7}
\end{align}
{and}
\begin{align}
\chi ^{\prime \prime }+\frac{2}{r}\chi ^{\prime }& =V_{\chi }.  \label{mov8}
\end{align}

In this way, let us impose that the potential $V(\phi ,\chi )$ can be represented in terms of a
position dependent formula and a  superpotential $W(\phi ,\chi )$ as%
\begin{equation}
V(\phi ,\chi )=\frac{1}{2r^{4}}\left[ \left( \frac{\partial W(\phi ,\chi )}{%
\partial \phi }\right) ^{2}+\left( \frac{\partial W(\phi ,\chi )}{\partial
\chi }\right) ^{2}\right] .  \label{2}
\end{equation}
This proposed form to relate the superpotential  to the scalar field potential is  one of the key dynamical assumptions to
allow analytical solution of the gravity and field equations. Note, that the  factor $1/r^4$ enhances the strength of the potential towards the 
galaxy center, which can be a natural assumption if the center of the galaxy has a particular property and it encompasses a more 
complex dynamical situation with the self-interaction of the DM field and its structure. For example, the increase of the effective interaction strength can be associate with fields of more complex structure, e.g., more components, vector/tensor  fields and group structure.  
Of course, such speculative assumption of our model can only be substantiated by the results we will show for the galaxy rotation curves.
However, it seems unprobable that the interaction of the DM with visible 
matter is the source of such enhancement, as if this direct interaction beyond gravity exists it should be  much weaker than the weak force, and even  on the galaxy scenario it is unlikely that it could eventually make some difference to justify the enhancement factor of the self interaction towards the center of the galaxies. On the other side, going towards the  galaxy border the DM self interactions tend to vanish and results in a non self-interacting DM scenario. 
Important to say that, the particular enhancement factor $1/r^4$ was chosen to allow analytical solutions of the field equations in the Newtonian gravity scenario.

Thus,  using the above representation, the following set of first-order differential equations are those that satisfy Eqs. (\ref{mov7}) and (\ref{mov8})
\begin{equation}
\frac{d\phi }{dr}=\frac{1}{r^2}\frac{\partial W(\phi ,\chi )}{\partial
\phi } ~~\mathrm{and}~~\frac{d\chi }{dr}=\frac{1}{r^2}\frac{\partial W(\phi ,\chi )}{\partial \chi }.
\label{13.1}
\end{equation}

It is possible from the above equation to formally
write the equation%
\begin{equation}
\frac{d\phi }{W_{\phi }}=\frac{dr}{r^2}=\frac{d\chi }{W_{\chi }},  \label{3.2}
\end{equation}

\noindent which leads to \cite{DutraPLB}%
\begin{equation}
\frac{d\phi }{d\chi }=\frac{W_{\phi }}{W_{\chi }}.  \label{3.3}
\end{equation}

It is worth to note that, the above equation is a nonlinear differential equation relating the scalar
fields $\phi$ and $\chi$ of the model so that $\phi =\phi (\chi )$. Then, once this function is known, Equations (\ref{13.1}) become uncoupled and can be solved.

As one can see, using this approach, solutions of the second-order differential equations (\ref{mov7}) and (\ref{mov8}) can be {obtained} through the corresponding first-order differential equations. In the next section, from the above equations, we will study a  consistent nonlinear model which has analytical solutions.

\section{4. Analytical Model}

In order to work with analytical solutions, we  will present the models proposed  in \cite{Bazeia,Bazeia2} and used for modeling a great number of systems \cite{b1,b2,b3,b4}, whose superpotential is given by

\begin{equation}
W(\phi ,\chi )=-\lambda \phi +\frac{\lambda }{3}\phi ^{3}+\mu \phi \chi ^{2},
\label{4w}
\end{equation}

\noindent where $\lambda $ and $\mu $ are real and positive dimensionless
coupling constants. Note that the $\phi$ and $\chi$ fields are divided by some arbitrary mass,
which will be not relevant to our solutions and in the final stage to obtain the rotation curve the full units will be recovered. 

As {pointed out in} \cite{DutraPLB}, general solutions of the first-order
differential equations can be found for the scalar fields, by first integrating the relation

\begin{equation}
\frac{d\phi }{d\chi }=\frac{W_{\phi }}{W_{\chi }}=\frac{\lambda (\phi
^{2}-1)+\mu \chi ^{2}}{2\mu \phi \chi }  \label{5}
\end{equation}%
and then by rewriting one of the fields in terms of the other.

Now, it is necessary to introduce the new variable $\rho =\phi ^{2}-1$. Then, we can rewrite the above
equation as

\begin{equation}
\frac{d\rho }{d\chi }-\frac{\lambda \rho }{\mu \chi }=\chi . \label{6}
\end{equation}

\noindent Solving the above equation, we obtain the corresponding general solutions

\begin{eqnarray}
\rho (\chi ) &=&\phi ^{2}-1=c_{0}\chi ^{\lambda /\mu }-\frac{\mu }{\lambda
-2\mu }\chi ^{2},\text{ }(\lambda \neq 2\mu ),  \label{7} \end{eqnarray}%
{}
\begin{eqnarray}
\rho (\chi ) &=&\phi ^{2}-1=\chi ^{2}[\ln (\chi )+c_{1}],\text{ }(\lambda
=2\mu ),  \label{8}
\end{eqnarray}%
where $c_{0}$ and $c_{1}$ are arbitrary integration constants. Substituting
the above solutions in the first-order differential equation for the field $%
\chi $, we have

\begin{eqnarray}
\frac{d\chi }{dr} &=&\pm 2\mu \chi \sqrt{1+c_{0}\chi ^{\lambda /\mu }-\frac{%
\mu }{\lambda -2\mu }\chi ^{2}},(\lambda \neq 2\mu )  \label{9}
\end{eqnarray}
{and}
\begin{eqnarray}
\frac{d\chi }{dr} &=&\pm 2\mu \chi \sqrt{1+\chi ^{2}[\ln (\chi )+c_{1}]},%
\text{{\ \ \ }}{(}\lambda =2\mu ).  \label{10}
\end{eqnarray}

It was shown in \cite{ DutraPLB} that there are four classes of analytical solutions for the present model, which we present below.

\subsection{Degenerate Solution Type-I} 

When $c_{0}<-2$ and $\lambda =\mu$ we have

\begin{eqnarray}
\chi _{DS}^{(1)}(r) &=&-\frac{2}{\left( \sqrt{c_{0}^{2}-4}\right) \cosh
(2\mu (r-r_0))-c_{0}}  \label{11} \\
&&  \notag 
\end{eqnarray}
and 
\begin{eqnarray}
\phi _{DS}^{(1)}(r) &=&-\frac{\left( \sqrt{c_{0}^{2}-4}\right) \sinh (2\mu (r-r_0))%
}{\left( \sqrt{c_{0}^{2}-4}\right) \cosh (2\mu (r-r_0))-c_{0}}.  \label{12}
\end{eqnarray}

\subsection{Degenerate Solution Type-II} 

For $\lambda =4\mu$
and $c_{0}<1/16$, the solutions are given by

\begin{eqnarray}
\chi _{DS}^{(2)}(r) &=&-\frac{2}{\sqrt{\left( \sqrt{1-16c_{0}}\right) \cosh
(4\mu (r-r_0))+1}}  \label{13} \\
&&  \notag 
\end{eqnarray}
{and}
\begin{eqnarray}
\phi _{DS}^{(2)}(r) &=&-\frac{\left( \sqrt{1-16c_{0}}\right) \sinh (4\mu (r-r_0))}{%
\left( \sqrt{1-16c_{0}}\right) \cosh (4\mu (r-r_0))+1}.  \label{14}
\end{eqnarray}

\subsection{Critical Solution Type-I}

For $\lambda =\mu$ and $c_{0}=-2$, we have the following set of solutions 

\begin{eqnarray}
\chi _{CS}^{(1)}(r) &=&\frac{1}{2}\left[ 1- \tanh (\mu (r-r_0))\right] 
\label{15} 
\end{eqnarray}
{and}
\begin{eqnarray}
\phi _{CS}^{(1)}(r) &=&-\frac{1}{2}\left[ \tanh [\mu (r-r_0))+ 1\right].
\label{16}
\end{eqnarray}

\subsection{Critical Solution Type-II}

Finally, when $\lambda =4\mu$ and $c_{0}=1/16$, the solution can be written as

\begin{eqnarray}
\chi _{CS}^{(2)}(r) &=&\sqrt{2}{\ }\frac{\cosh (\mu (r-r_0))+ \sinh (\mu (r-r_0))}{%
\sqrt{\cosh (2\mu (r-r_0))}}  \label{17} 
\end{eqnarray}
{and}
\begin{eqnarray}
\phi _{CS}^{(2)}(r) &=&-\frac{1}{2}\left[ 1+\tanh (2\mu (r-r_0))\right] .
\label{18}
\end{eqnarray}

\begin{figure}[h]
\includegraphics[width=8.3cm]{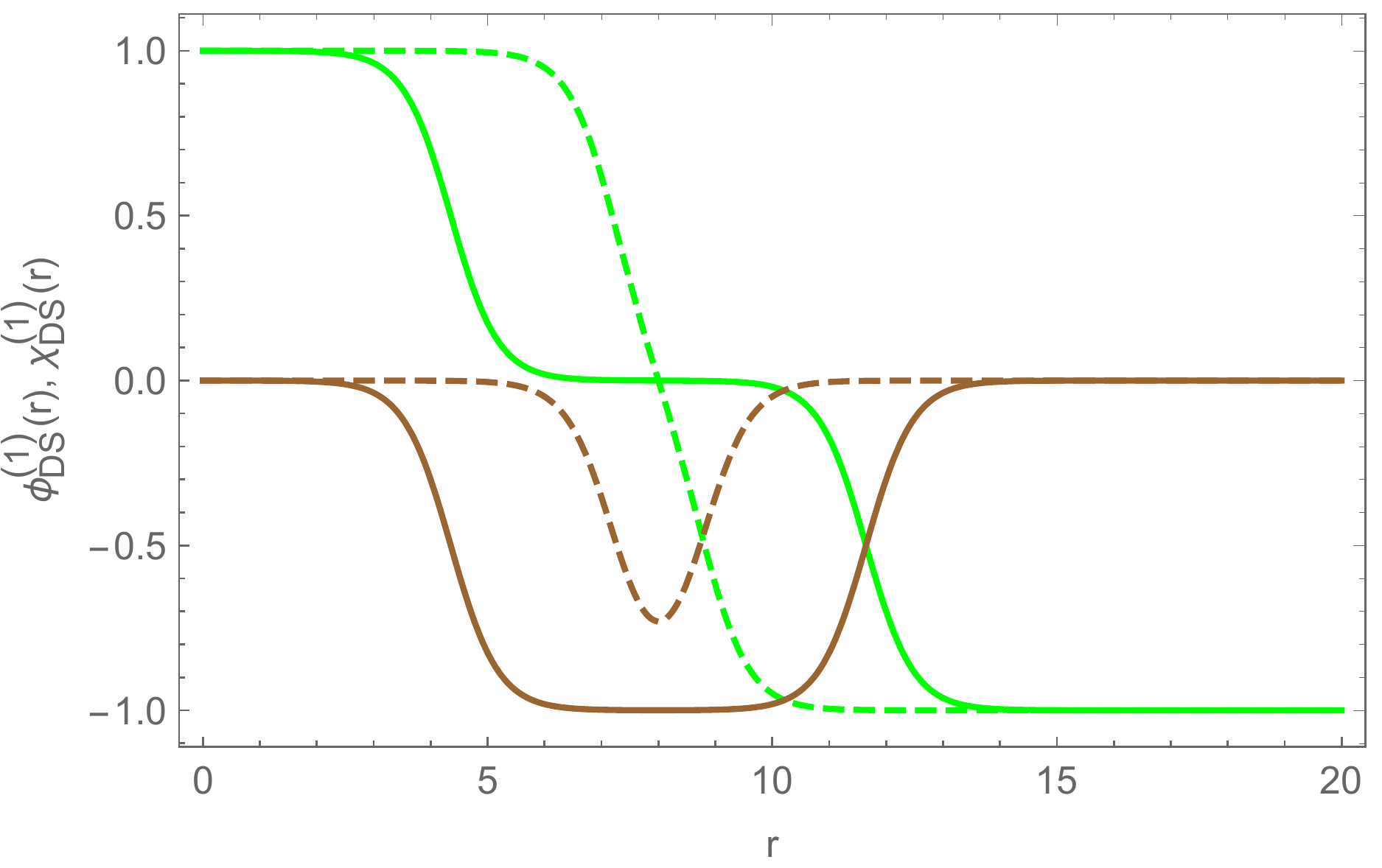}
\includegraphics[width=8.3cm]{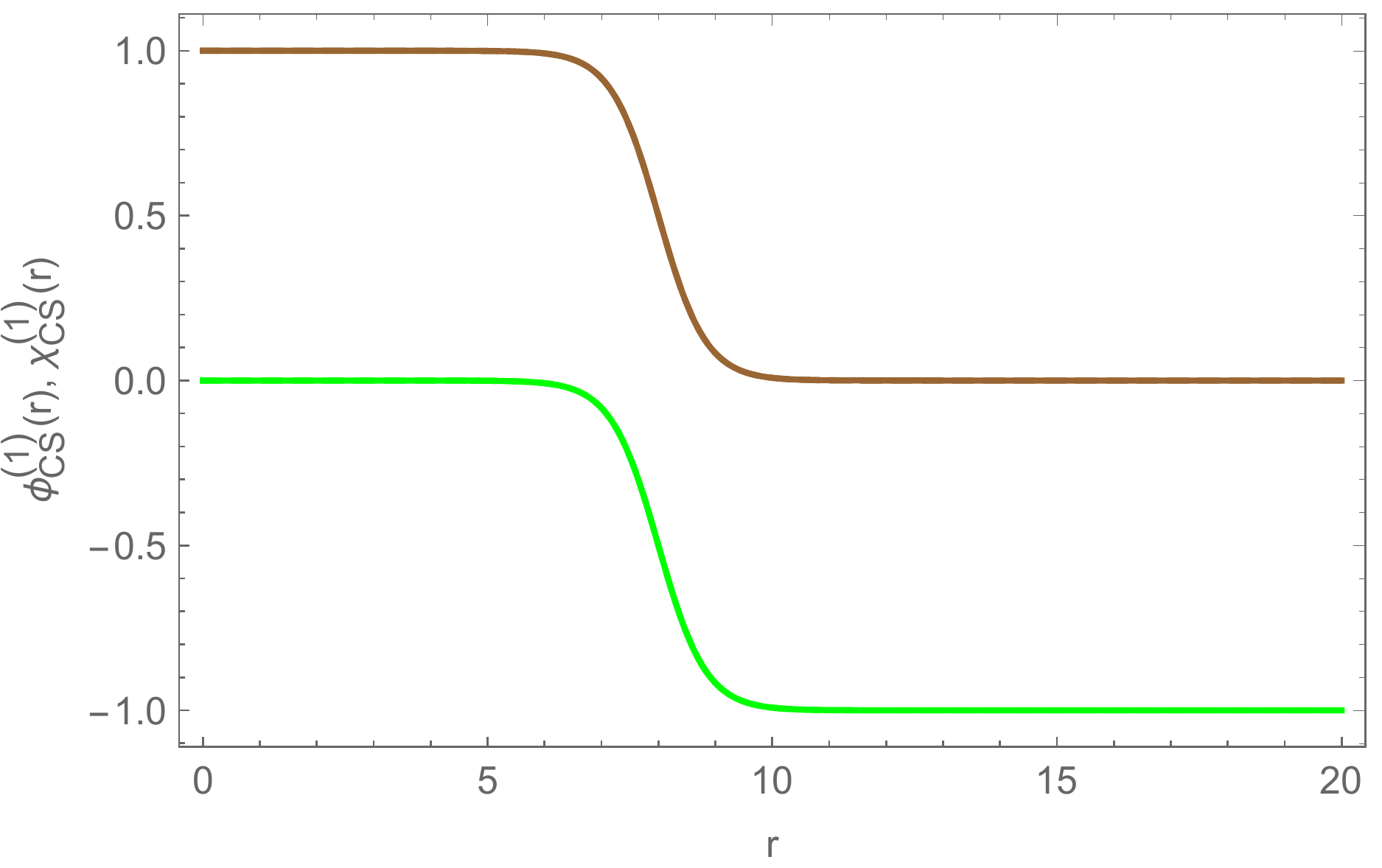}
\caption{Degenerate (top { panel} ) and critical (bottom {panel}) type-I solutions with $\mu=1.2$. Green lines correspond to the field $\phi$ {while brown lines} to the field $\chi$. In the top panel, dashed lines represent the cases with $c_0=-2.0000001$ and continuous lines  $c_0=-2.1$,  {\rm assuming} $r_0=8$. 
In the bottom panel, the value $r_0=0$ is assumed.}
\label{fig1}
\end{figure}

In Figure~\ref{fig1} above we show some typical profiles of the type-I degenerate and critical solutions. Note that both two-kink and flat-top lump solutions arise for values of $c_0$ close to the critical value of $-2$. We emphasize that  solutions of type-II have a similar behavior to the configurations of type-I.

{Subsequently},  we will use the above models to describe the profile of rotation curves of galaxies. In particular, { our  goal  is to find successful fittings of the models to the rotation curves obtained through observations of} 
 dwarf and low surface brightness (LSB) late-type galaxies \cite{kratsov}. 

\bigskip

\section{5. Rotation Curves}
In this section, we will study the rotation curves for galaxies in the presence of the models 
described in the previous section. Our aim is to show that it is possible to find a
robust theoretical fit for such curves. In view of this, let us begin by writing the equation that describes the general rotation curve of galaxies \cite{Mielke} 

\begin{equation}
    v^{2}(r) \simeq \frac{k}{2}\left(\frac{M(r)}{r}+p_{r}r^{2}\right), \label{vel}
\end{equation}
\noindent where $M(r)$ is the resulting mass function

\begin{equation}
    M(r)=\int _{0}^{r} dy y^{2} \rho(y). \label{m}
\end{equation}

Considering the Newtonian limit, we can rewrite the above equation as

\begin{eqnarray}
&&\left. M(r)=\int_{0}^{r}dyy^{2}\left\{ \frac{1}{2}\left[ \left( \phi
^{\prime }\mp \frac{W_{\phi }}{y^{2}}\right) ^{2}+\left( \chi ^{\prime }\mp 
\frac{W_{\chi }}{y^{2}}\right) ^{2}\right] \right. \right.   \notag \\
&&\left. \left. \mp \frac{W_{\phi }\phi ^{\prime }}{y^{2}}\mp \frac{W_{\chi
}\chi ^{\prime }}{y^{2}}\right\}.\right.   \label{m1}
\end{eqnarray}

{Then}, making use of the relation given by Eqs. \eqref{13.1} of the superpotential with the fields, one finds that
\begin{equation}
M(r)=\pm \{W[\phi (r),\chi (r)]-W[\phi (0),\chi (0)]\}.  \label{m2}
\end{equation}

It can be seen that by applying this approach the mass function depends only on the superpotential, thereby allowing us to calculate in a simple way the analytical expression for the rotation curve of a disc galaxy. 
Therefore, for the model under analyses, we obtain the following rotation velocities.

\begin{widetext}

\subsection{Rotation Velocity: Type-I Degenerate Solution}

\begin{eqnarray}
&&\left. v_{DS-I}^{2}(r)=\frac{1}{3}\frac{\sqrt{c_{0}^{2}-4}}{r}\left\{ \frac{%
\left( c_{0}^{2}-4\right) \lambda \sinh ^{3}(2\mu (r-r_{0}))}{\left( c_{0}-%
\sqrt{c_{0}^{2}-4}\cosh (2\mu (r-r_{0}))\right) ^{3}}+\sinh (2\mu
(r-r_{0}))\times \right. \right.   \notag \\
&&\left. \left[ \frac{3\lambda }{\sqrt{c_{0}^{2}-4}\cosh (2\mu
(r-r_{0}))-c_{0}}+\frac{12\mu }{\left( c_{0}-\sqrt{c_{0}^{2}-4}\cosh (2\mu
(r-r_{0}))\right) ^{3}}\right] \times \right.   \notag \\
&&\left. \left. \left[ \frac{\sinh (2\mu r_{0})\left( 5c_{0}^{2}\lambda -6%
\sqrt{c_{0}^{2}-4}c_{0}\lambda \cosh (2\mu r_{0})+\left( c_{0}^{2}-4\right)
\lambda \cosh (4\mu r_{0})-8\lambda -12\mu \right) }{\left( c_{0}-\sqrt{%
c_{0}^{2}-4}\cosh (2\mu c_{0})\right) ^{3}}\right] \right\} \right. .
\label{sol1}
\end{eqnarray}
\subsection{Rotation Velocity: Type-II Degenerate Solution}
\begin{eqnarray}
&&\left. v_{DS-II}^{2}(r)=\frac{1}{3}\frac{\sqrt{1-16c_{0}}}{r}\left\{ \frac{%
(16c_{0}-1)\lambda \sinh ^{3}(4\mu (r-r_{0}))}{\left( \sqrt{1-16c_{0}}\cosh
(4\mu (r-r_{0}))+1\right) ^{3}}\right. \right.   \notag \\
&&\left. +\frac{3\sinh (4\mu (r-r_{0}))\left[ \lambda \left( \sqrt{1-16c_{0}}%
\cosh (4\mu (r-r_{0}))+1\right) ^{2}-4\mu \right] }{\left( \sqrt{1-16c_{0}}%
\cosh (4\mu (r-r_{0}))+1\right) ^{3}}\right.   \notag \\
&&\left. \left. +\frac{\sinh (4\mu r_{0})\left( -32c_{0}\lambda +6\sqrt{%
1-16c_{0}}\lambda \cosh (4\mu r_{0})+(\lambda -16c_{0}\lambda )\cosh (8\mu
r_{0})+5\lambda -12\mu \right) }{\left( \sqrt{1-16c_{0}}\cosh (4\mu
r_{0})+1\right) ^{3}}\right\} \right. .  \label{sol2}
\end{eqnarray}
\subsection{Rotation Velocity: Type-I Critical Solution}
\begin{eqnarray}
v_{CS-I}^{2}(r) &=&\frac{1}{24r}\left\{ \tanh (\mu (r-r_{0}))[3(3\lambda
+\mu )+\tanh (\mu (r-r_{0}))(-3\lambda +3\mu -(\lambda +3\mu )\tanh (\mu
(r-r_{0})))]\right.   \notag \\
&&\left. -(\lambda +3\mu )\tanh ^{3}(\mu r_{0})+3(\lambda -\mu )\tanh
^{2}(\mu r_{0})+3(3\lambda +\mu )\tanh (\mu r_{0})\right\}   \label{sol3}
\end{eqnarray}
\subsection{Rotation Velocity: Type-II Critical Solution}
\begin{eqnarray}
v_{CS-II}^{2}(r) &=&\left\{ -\lambda \lbrack \tanh (2\mu
(r-r_{0}))-1]^{3}+12\lambda \lbrack \tanh (2\mu (r-r_{0}))-1]+24\mu \text{%
sech}^{2}(2\mu (r-r_{0}))\right.   \notag \\
&&\left. -\lambda \lbrack \tanh (2\mu r_{0})+1]^{3}+12\lambda \lbrack \tanh
(2\mu r_{0})+1]-24\mu \text{sech}^{2}(2\mu r_{0})\right\} .  \label{mov4}
\end{eqnarray}
\end{widetext}

\begin{figure}[h]
\includegraphics[width=8.2cm]{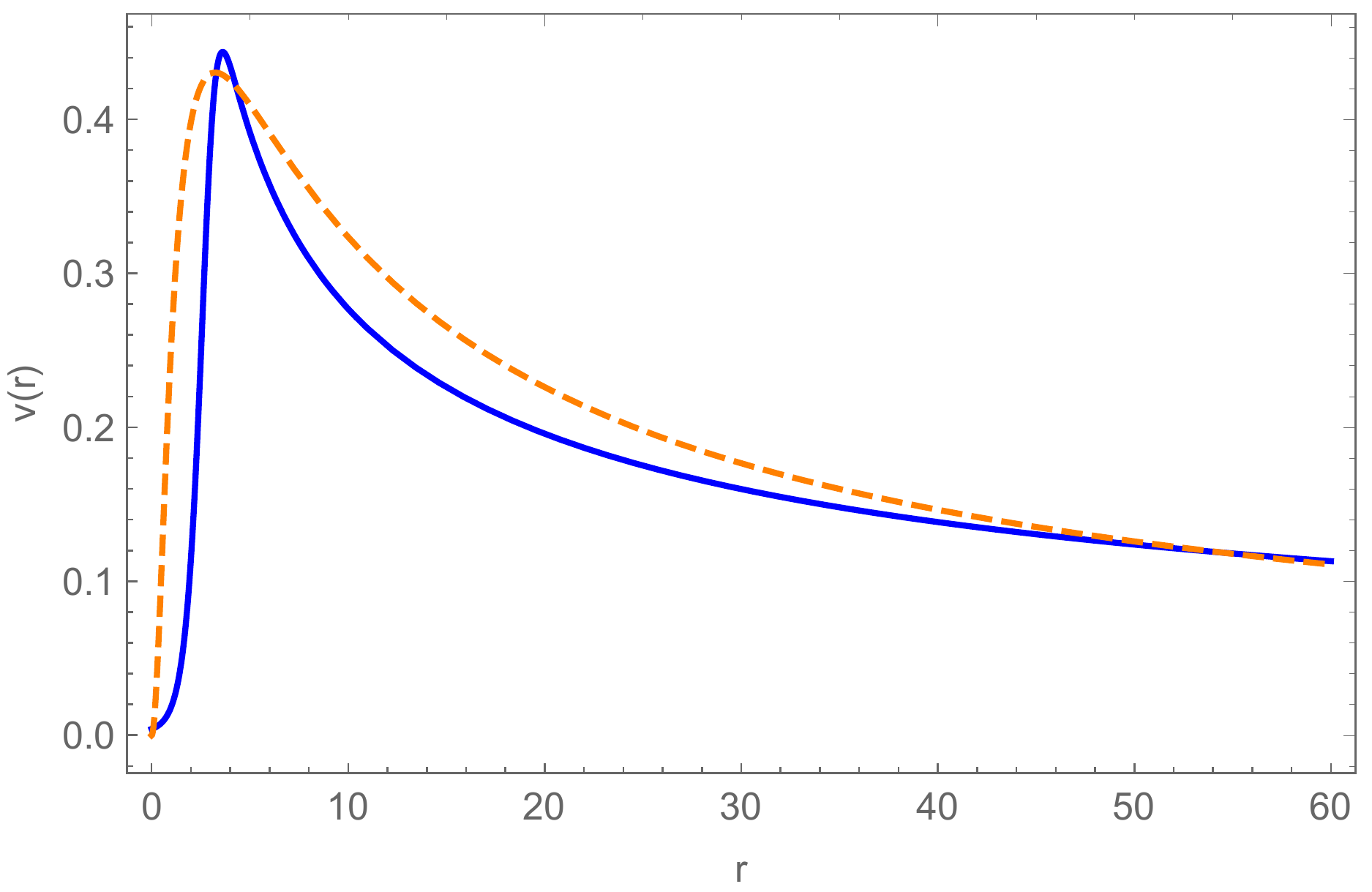}
\includegraphics[width=8.2cm]{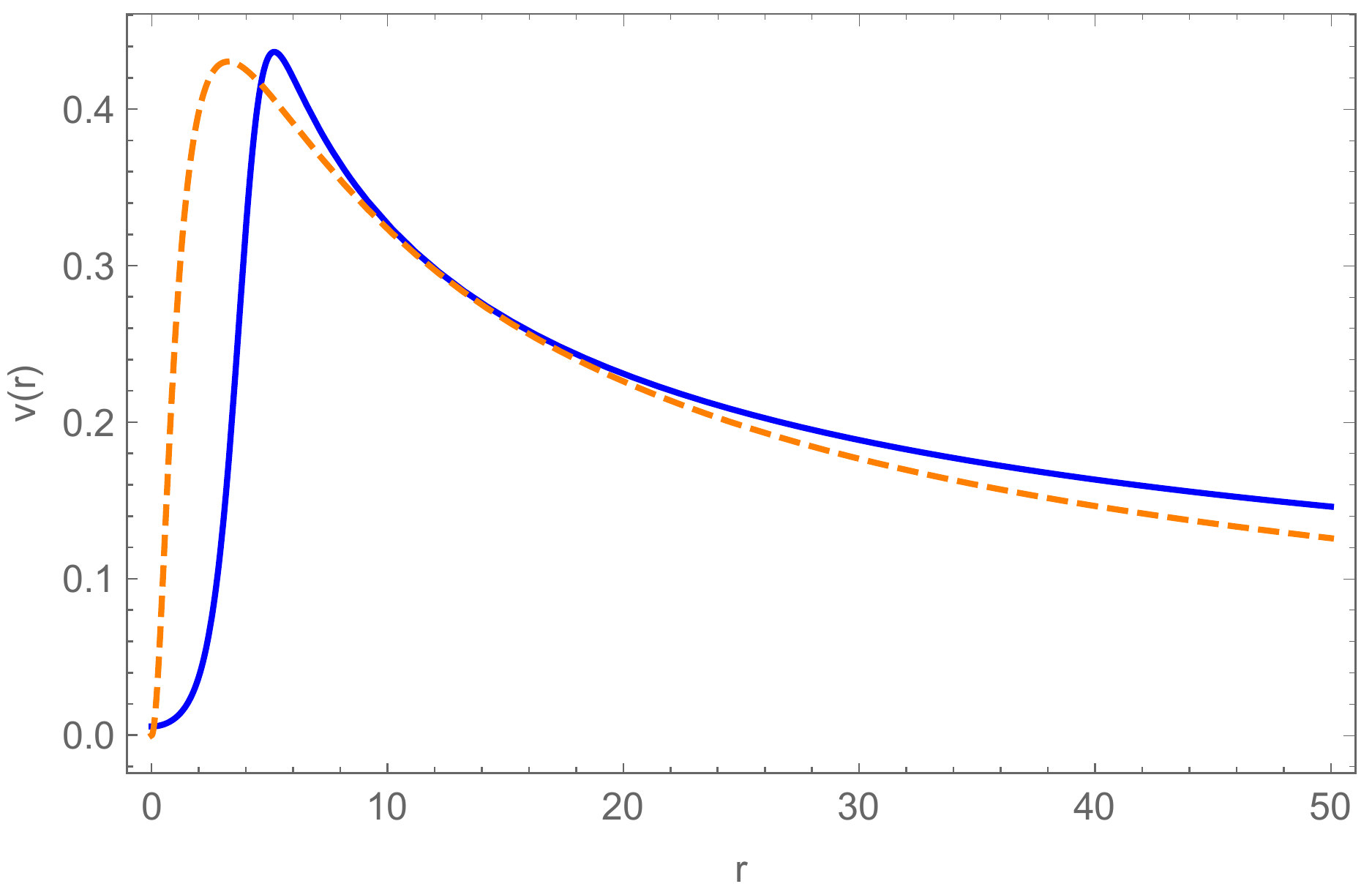}
\caption{Rotation curves in dimensionless units. Blue continuous line are the solutions for degenerate solution and orange dashed lines are the observational fit showed by Bukert \cite{bukert}. Top figure is the rotation curve for the type-I degenerate solution with $c_0=-2.0000001$, $\mu=1.4$ and $r_0=-0.9$. On the other hand, the bottom figure is the degenerate type-II solution, with $c_0=1/16.0001$, $\mu=0.4$ and $r_0=0$.}
\label{fig2}
\end{figure}

\begin{figure}[h]
\includegraphics[width=8.2cm]{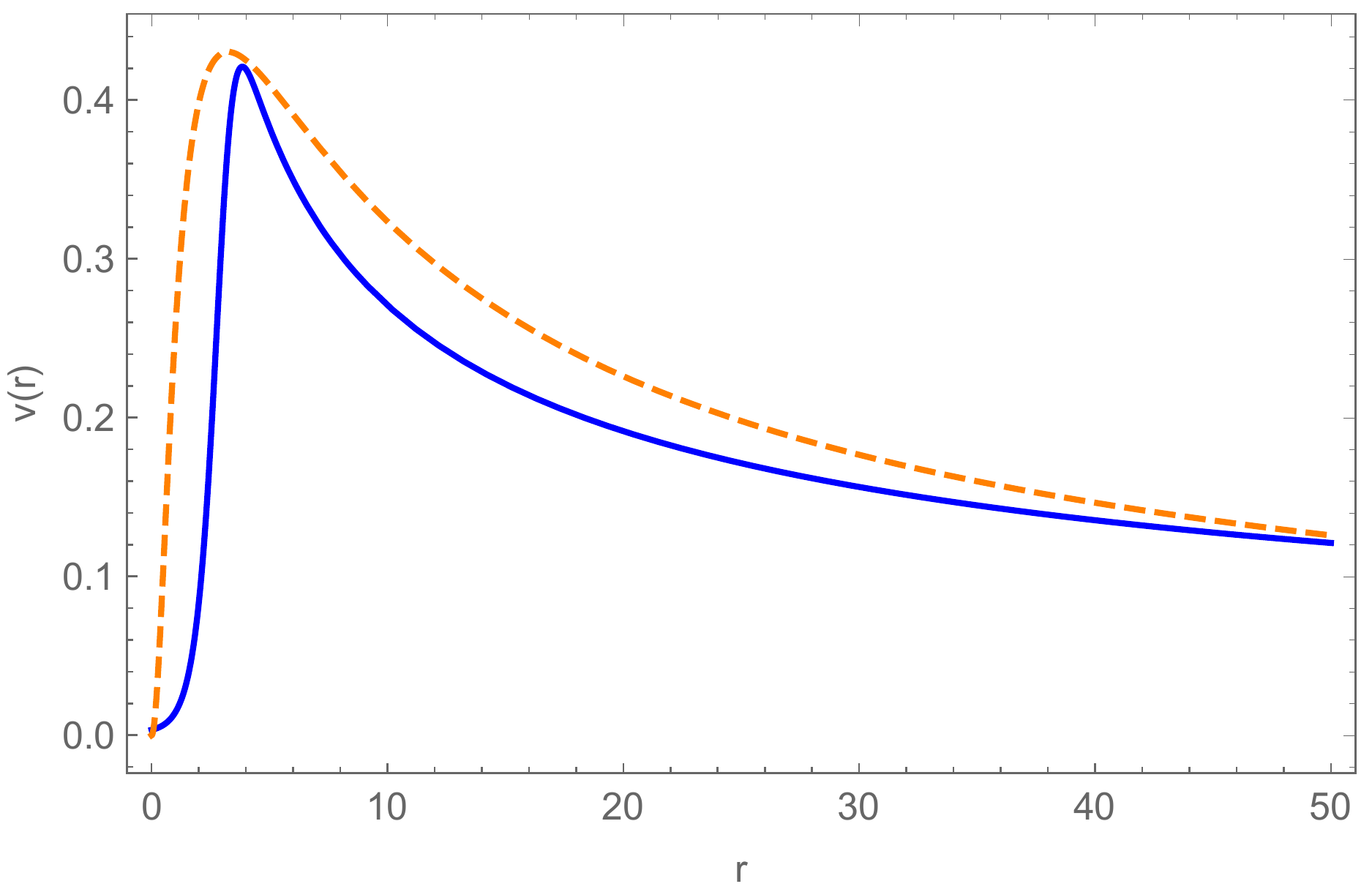}
\includegraphics[width=8.2cm]{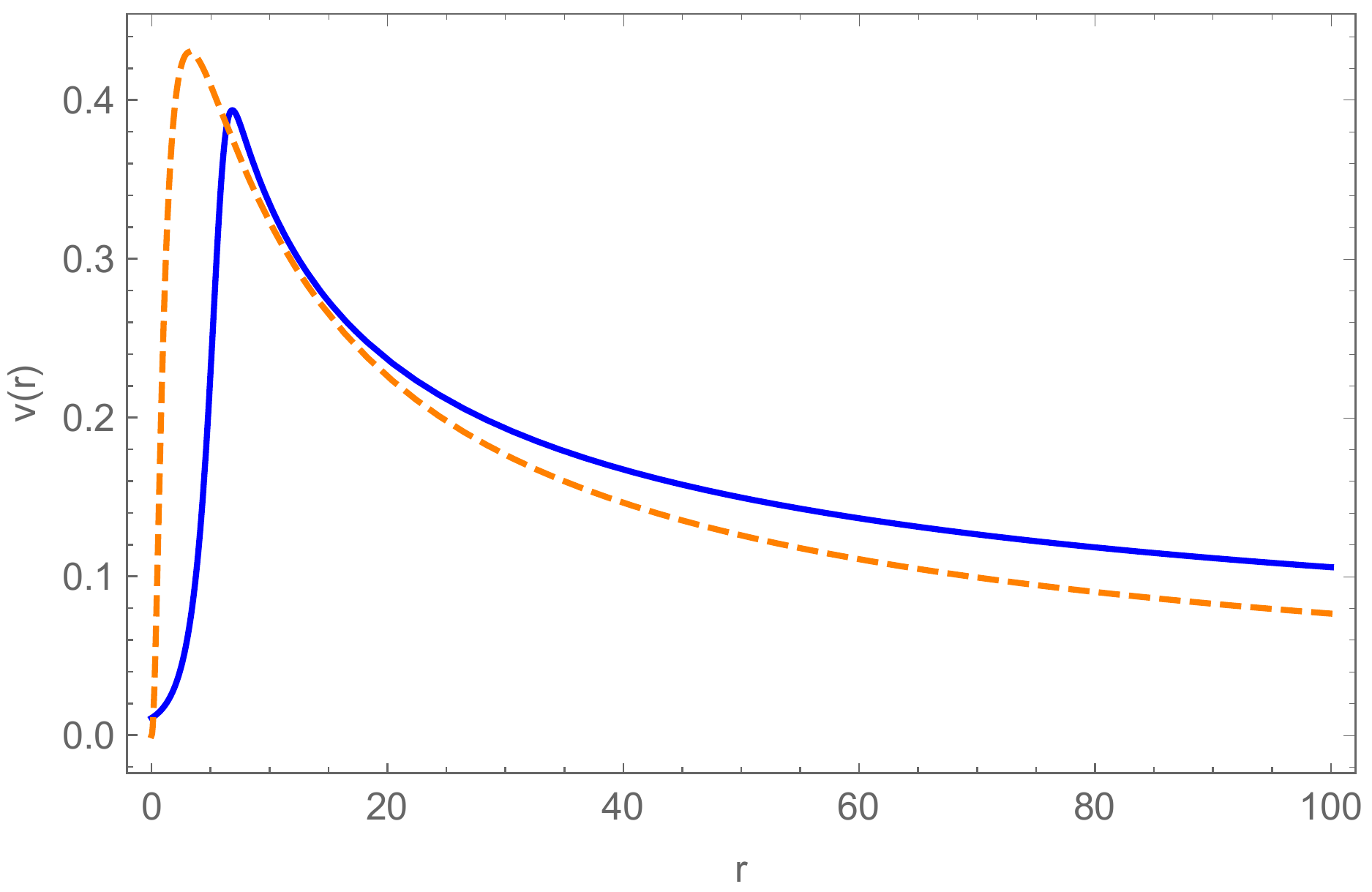}
\caption{Rotation curves in dimensionless units. Blue continuous line are the solutions for critical solutions and orange dashed lines are the observational fit showed by Bukert \cite{bukert}. Top figure is the rotation curve for the type-I critical solution with $\mu=1.1$ and $r_0=3.1$. On the other hand, the bottom figure is the critical solution type-II, with $\mu=0.42$ and $r_0=6$.}
\label{fig3}
\end{figure}

The profiles of the rotation curves from these model are shown in Figs. \ref{fig2} and \ref{fig3}. From those figures, we see that the observational rotation curves taken from \cite{bukert} can be fitted by our analytical solutions.

\section{6. Conclusions}

DM is one of the greatest mysteries of Physics. Nowadays there is a plethora of possibilities to model DM (check also  \cite{bertone/2005,cheng/2002,petraki/2013,seljak/2006}). The difficulty in understanding the DM nature has even led to some attempts to substitute it by purely geometrical effects coming from extensions of General Theory of Relativity (besides \cite{deliduman/2020,o'brien/2018,mak/2004,capozziello/2004}, check also \cite{mannheim/2013,obrien/2012,obrien/2018,capozziello/2007}).  

In the present work, we have adopted the BEC DM scenario of a complex scalar field coupled to gravity.
We have shown that splitting the complex scalar field, which is responsible for the nucleation of the bosonic condensate, in its real and imaginary parts allows us to map the problem in an effective theory with two fields. Here, in the Newtonian approach, we developed an analogous technique  to the orbit procedure \cite{DutraPLB}, where the second-order field equations were reduced to a pair of coupled first-order equations.

By analyzing the model given by Eqs.~(\ref{13.1}) and the superpotential from Eq.~\eqref{4w}, we presented a rich class of analytical solutions for the scalar fields, which describe reasonably the observational fit proposed by Burkert \cite{bukert} for the DM halos of dwarf spiral galaxies. Moreover, since Burkert empirical fitting formula is nearly identical to rotation curves of a sample of DM dominated dwarf and LSB late-type galaxies \cite{kratsov}, our analytical solutions are  also well matched with those observational results.

Important to point out that the key dynamical assumption to
allow analytical solution of the gravity and field equations, is the assumed form of the relation between the superpotential and the field potential,
which presents a position dependent relation with an enhancement of the self-interaction of the scalar fields towards the galaxy center. On the other side, 
 going towards the  galaxy border  the interaction tends to vanish building a non self-interacting DM scenario. The dependence of the effective interaction strength presumably should be originated from a  more complex structure of the fields, e.g., more components, vector/tensor  fields and group structure.  
Our speculative assumptions were substantiated by the reasonable reproduction of the galaxy rotation curves. We deem that it is unlikely 
that the interaction of the DM with visible matter is the source of such position dependent strength, as even if this direct interaction beyond gravity exists it should be  much weaker than the weak force, and  on the galaxy density scenario it is unlikely that it could eventually make some difference to justify the enhancement factor of the self interaction towards the center of the galaxies.

We stress here that our approach presents a good fit to the analytical observational curve even for larger values of $r$ (check, for instance, \cite{Mielke}), while for small radius the free parameters of our model can deal with eventual discrepancies. Furthermore, comparing to previous studies within the
BEC scenario of DM galaxy halos, our results are based on a fully analytical solution of a consistent dynamical nonlinear approach 
within the Newtonian gravity and its BEC matter source.
As we can see, the approach shown in our work is general, thus paving the way to investigate new theoretical models in DM scenarios.

\bigskip

\textbf{Acknowledgements}

RACC thanks to S\~ao Paulo Research Foundation (FAPESP), grant numbers 2016/03276-5 and 2017/26646-5 for financial support. RACC also thanks Prof. Elisa G. M. Ferreira for introducing him this matter and for valuable discussions and comments. Moreover, RACC thanks Prof. Eiichiro Komatsu for the opportunity to visit the Max Planck Institute for Astrophysics (MPA), where this work was started. PHRSM thanks CAPES for financial support. OLD thanks to FAPESP and CNPq. ASD, TF and WP thanks CAPES, CNPq, and FAPESP for financial support.

\end{document}